\documentstyle[11pt,newpasp,twoside,epsf]{article}
\def\edcomment#1{\iffalse\marginpar{\raggedright\sl#1\/}\else\relax\fi}
\reversemarginpar

\begin{document}
\title{A Comparative Study of Damped Ly$\alpha$ Galaxies}
\author{David A. Turnshek, Sandhya M. Rao, and Daniel B. Nestor}
\affil{Department of Physics \& Astronomy, University of Pittsburgh, Pittsburgh,
PA, 15260, USA}


\begin{abstract}
We compare the properties of a sample of local ($z=0$) gas-rich
galaxies studied in 21 cm emission to a sample of 13 low-redshift
($z<1$) damped Ly$\alpha$ (DLA) galaxies identified as the counterparts
of low-redshift DLA systems found in QSO absorption-line surveys.
This absorption-selected sample has average redshift $<\!z\!> = 
0.5$.  We find that many of the properties of the two samples are
comparable. However, consideration of the statistical results on 
all known low-redshift DLA systems indicates that there is: (1) a somewhat
higher incidence and cosmological mass density for the low-redshift DLA
systems in comparison to expectations at $z=0$ and (2) an unexpectedly
high rate of occurrence of very large column density ($N_{HI} > 10^{21}$
atoms cm$^{-2}$) low-redshift DLA systems; both of these results are
discussed by Rao \& Turnshek in these proceedings. These differences,
coupled with imaging studies, suggest that there may be an excess
of low-redshift DLA galaxies in the form of dwarf and/or low surface
brightness galaxies. Some examples of low-redshift DLA galaxies are
shown by Nestor et al. in these proceedings.

\end{abstract}

\section{Introduction}

One of the future challenges in the field of galaxy formation research
will be to {\it self-consistently} determine and understand the redshift
evolution of the various global parameters which describe the conversion
of gas into stars on a cosmological scale: galaxy number counts as a
function of color, the (ionized, atomic, molecular) gas masses, star
formation rates, and metallicities including dust content. However, in
order to accomplish self-consistency, it will be necessary to determine
many of these parameters for a common set of objects, so that the biases
introduced by any single method used to take a census are clear. Here
we discuss work along these lines with the aim of studying the galaxies
associated with the bulk of the {\it neutral hydrogen} component in the
Universe. As previous studies have shown, this component can be traced
through the study of damped Ly$\alpha$ (DLA) systems in QSO spectra.

With the discovery of a moderate number of low-redshift ($z<1.65$) DLA
systems facilitated by $HST$ UV spectroscopy (Rao \& Turnshek 2000),
it is now possible to undertake meaningful studies of the galaxies
associated with these systems (e.g. Turnshek et al. 2001 and references
therein). Some of the results of this work can be summarized as follows:
(1) The low-redshift cosmological mass density of neutral gas contributed
by DLA systems, and to a lesser extent their incidence, is higher than
was expected, although the errors are still large. (2) The observed HI
column density distribution of the low-redshift DLA systems has an excess
of higher column density systems in comparison to what would be expected
if the column densities were derived from a sample of randomly-oriented
local gas-rich (disk) galaxies. (3) The low-redshift DLA galaxies are not
limited to luminous disk galaxies, but rather they include a significant
number of dwarf ($L < 0.1L^*$) and low surface brightness galaxies.

In these proceedings, Rao \& Turnshek (2001) and Nestor et al. (2001)
have reviewed these and other results on the evolution of DLA systems
and presented images of associated DLA galaxies, respectively.  Thus,
our remaining aim for these proceedings is to review in more detail
how the properties of identified low-redshift DLA galaxies compare
with local galaxies.  In \S2 the data and methods we use to assess the
properties of local galaxies and low-redshift DLA galaxies are summarized.
In \S3 we perform a comparative study of some of their
properties. In \S4 we discuss the conclusions that can be drawn from
the comparisons.  When needed, cosmological parameters of H$_0 = 65$
km s$^{-1}$, q$_0 = 0.5$, and $\Lambda = 0$ are adopted for ease of
comparison with published results.

\begin{figure} 
\plotfiddle{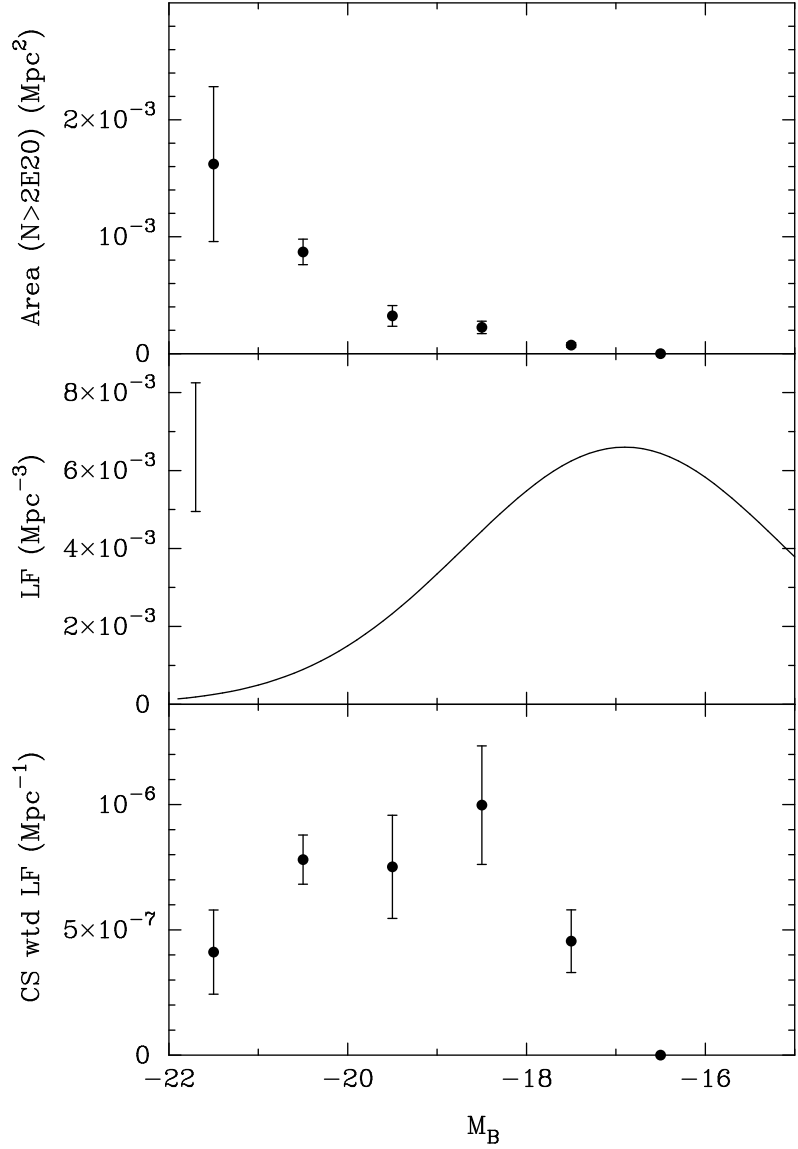}{15.5cm}{0}{70}{70}{-200}{-30}
\caption{\footnotesize The results shown above were derived from 21
cm emission observations of 27 local gas-rich galaxies originally
studied by Rao \& Briggs (1993) to determine the equivalent
$z=0$ incidence and cosmological mass density of DLA absorbers. For
random orientations, the top panel shows the sample's effective
cross-section as a function of luminosity (M$_{\rm B}$) for $N_{HI}
\ge 2\times10^{20}$ atoms cm$^{-2}$.  The middle panel shows the
luminosity function of Tammann (1986), which was adopted in the Rao
\& Briggs (1993) study. The vertical error bar in the upper left of
the middle panel represents the uncertainty in the luminosity
function's normalization at its peak, $\approx 25$\%.
The bottom panel shows the cross-section
weighted luminosity function of the local sample, which represents
the relative probability that a local ($z=0$) galaxy of a given
luminosity (M$_{\rm B}$) will give rise to a DLA system. This lower
panel is re-plotted in Figure 4 for comparison with low-redshift
($<\!z\!> = 0.5$) DLA galaxies.} \end{figure}

\section{Data}

\subsection{Local Galaxies}

Since the identification of a classical DLA absorption line requires
an HI column density $N_{HI} \ge 2\times10^{20}$ atoms cm$^{-2}$, we
are naturally interested in those local ($z=0$) galaxies which have
substantial neutral gas components. In particular, we are interested
in the properties of those local galaxies
that would give rise to classical DLA systems if they were observed
in front of background QSOs.  Thus, to perform a comparative
study between local galaxies and low-redshift DLA galaxies, it is
necessary to define a sample of local galaxies with well-studied
neutral hydrogen properties.

As our starting point for the comparative analysis, we used the
sample of 27 local galaxies studied by Rao \& Briggs (1993). Their
sample was selected from Nilson's (1973) Uppsala General Catalog of
Galaxies (i.e. the UGC).  The UGC reaches a limiting diameter
of $\approx 1$ arcmin and/or a limiting apparent magnitude of
$\approx 14.5$ mag on the blue Palomar Observatory Sky Survey
plates. Coverage on the sky is limited to north of declination
$-2.5$ deg.  The study by Cornell et al. (1987) has shown that the
B magnitude surface brightness limits used to estimate the galaxy
major-axis diameters in the UGC vary considerably, but mostly lie
between 25.0 and 26.5 mag arcsec$^{-2}$.  Rao \& Briggs (1993)
considered all objects in the UGC catalog larger than 7 arcmin on
the sky within the Arecibo declination range with velocities $> 200$
km s$^{-1}$. The velocity constraint was used to exclude members
of the Local Group.

The program to obtain 21 cm observations of the 27 local galaxies
was initiated by Briggs et al. (1980) and
later analysis of these observations was used to estimate the
equivalent incidence and cosmological mass density of DLA systems
at $z=0$ (Rao \& Briggs 1993); the errors in these determinations 
were later slightly
revised (Rao, Turnshek, \& Briggs 1995). For random orientations,
we have determined this sample's cross-section as a function of
galaxy luminosity (M$_{\rm B}$) for $N_{HI} \ge 2\times10^{20}$
atoms cm$^{-2}$ (upper panel of Figure 1).  Then, using the local
luminosity function of Tammann (1986, middle panel of Figure 1), which
was adopted by Rao \& Briggs (1993), we have derived a cross-section
weighted luminosity function for local ($z=0$) DLA galaxies 
(lower panel of Figure 1); this represents the relative probability
that a local galaxy of a given luminosity will give rise to a
classical DLA line. Our result is similar to a result derived 
from 21 cm emission observations of the Ursa Major cluster and 
presented by Zwaan at this workshop.
We also show the distribution of radii corresponding
to $N_{HI} \ge 2\times10^{20}$ atoms cm$^{-2}$ (Figure 2) for the
$z=0$ sample when viewed face-on; only 24 of the 27 galaxies have
regions with $N_{HI} \ge 2\times10^{20}$ atoms cm$^{-2}$ when viewed
face-on. The upper envelope of this distribution gives an indication
of the expected near-maximum value of the impact parameter as a
function of luminosity for local ($z=0$) DLA galaxies.

\begin{figure}[h] 
\plotfiddle{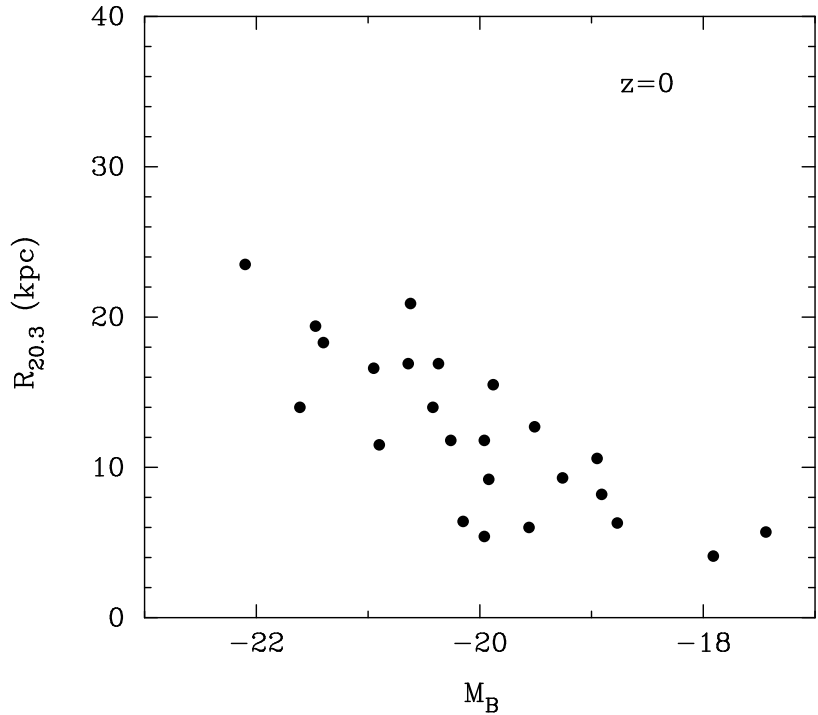}{10.2cm}{0}{75}{75}{-230}{-210}
\caption{\footnotesize Distribution of face-on radii corresponding
to $N_{HI} \ge 2\times10^{20}$ atoms cm$^{-2}$ ($R_{20.3}$)
as a function of luminosity (M$_{\rm B}$) for the local ($z=0$)
galaxy sample of Rao \& Briggs (1993).  Only 24 of the 27 galaxies
in their sample have $N_{HI} \ge 2\times10^{20}$ atoms cm$^{-2}$
when corrected to a face-on orientation. We note that the relative
number distribution of these points is not representative of the true
local number distribution, since there was no weighting by the local
luminosity function. However, since the results shown correspond to
face-on galaxy properties, the upper envelope of this distribution
gives an indication of the expected near-maximum value of a DLA
galaxy's impact parameter (b) as a function of galaxy luminosity
(M$_B$) at $z=0$.  This is because the probability of producing
a DLA line when $b > R_{20.3}$ is relatively small (but not zero)
for inclined galaxies. These points are re-plotted in Figure 5 for
comparison with low-redshift ($<\!z\!> = 0.5$) DLA galaxies.}
\end{figure}

\newpage

\subsection{Low-Redshift DLA Galaxies}

\begin{figure}[t] 
\plotfiddle{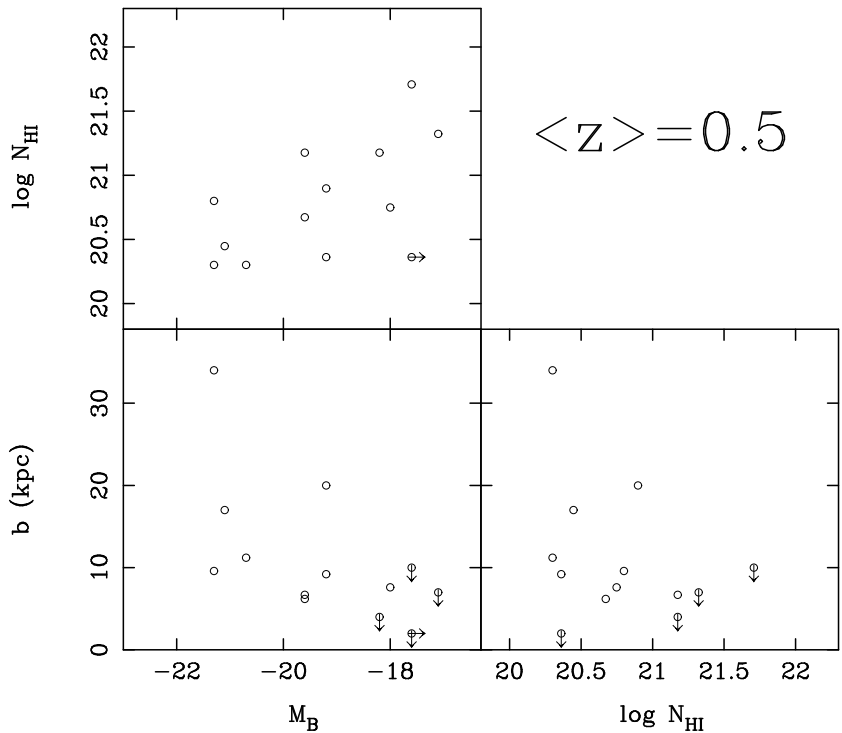}{11.0cm}{0}{80}{80}{-220}{-200}
\caption{\footnotesize The 3-dimensional distribution of luminosity
(M$_{\rm B}$), impact parameter ($b$), and neutral hydrogen
column density ($N_{HI}$) for our sample of 13 low-redshift
($z<1$) DLA galaxies with average redshift $<\!z\!> = 0.5$.
As discussed in \S2.2, a variety of methods were used to make DLA
galaxy ``identifications'' and it is difficult to rule out the
possibility that a fainter galaxy or a galaxy with a smaller impact
parameter is the actual DLA absorber.} \end{figure}

At the present time there are $\approx 12$ QSO fields with
$\approx 13$ low-redshift ($z<1$) DLA systems that have been
imaged with the goal of identifying the associated DLA galaxies
and studying their properties. The average redshift of this sample is
$<\!z\!> = 0.5$.  It should be kept in mind that the degree of
confidence for any ``identification'' of a DLA galaxy is variable,
with the confidence being highest when the candidate DLA galaxy
has a low impact parameter along the QSO sight-line {\it and}
there is a confirming slit spectrum showing that it is at the DLA
system redshift.  However, even when the confidence is relatively
high, it is possible that some other galaxy, possibly a fainter
galaxy or a galaxy with a smaller impact parameter, might be the
actual DLA absorber.

As a rule, three methods have been used to make a DLA galaxy
identification: (1) a slit redshift, (2) a photometric redshift,
and/or (3) simply the existence of an isolated galaxy at relatively
low impact parameter along the QSO sight-line. See Turnshek
et al. (2001) and Nestor et al. (2001) for examples of the first
two methods and Le Brun et al. (1997) for examples of the third
method. The results of the identifications for 13 low-redshift
DLA galaxies are shown in Figure 3. Provided in this figure is
the 3-dimensional distribution of DLA galaxy luminosity (M$_{\rm
B}$), impact parameter ($b$), and neutral hydrogen column density
($N_{HI}$) for the 13 identifications.

\begin{figure}[h]
\plotfiddle{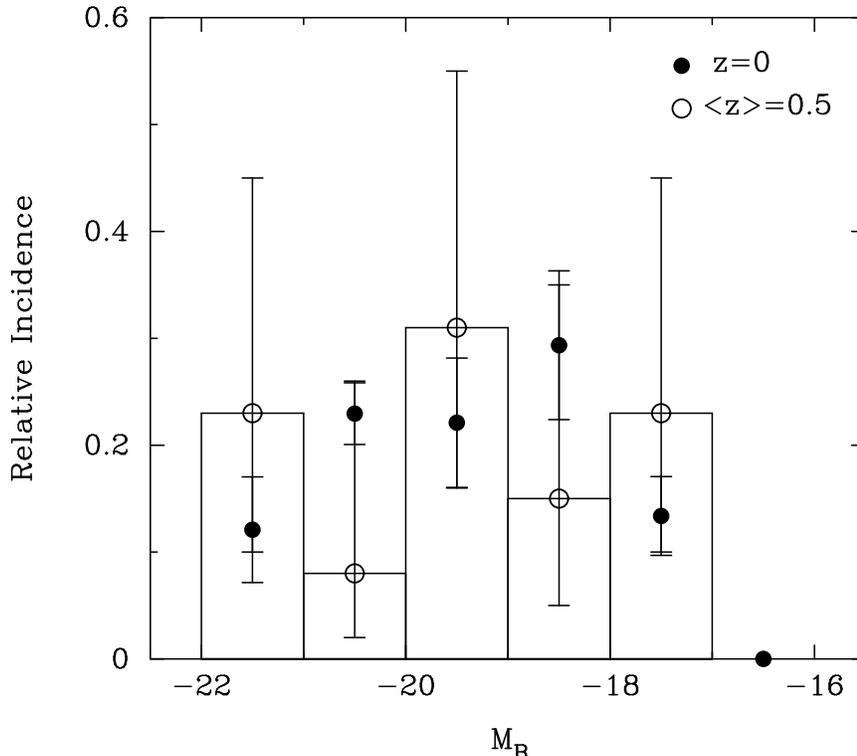}{9.3cm}{0}{80}{80}{-220}{-230}
\caption{\footnotesize Relative luminosity (M$_{\rm B}$)
distributions of low-redshift ($<\!z\!> = 0.5$) DLA galaxies in
comparison to the derived cross-section weighted luminosity function
of local ($z=0$) DLA galaxies (i.e. for $N_{HI} \ge 2\times 10^{20}$
atoms cm$^{-2}$) which was shown in Figure 1. The two normalized
distributions are roughly comparable.} \end{figure}

\section{Comparison of $z=0$ and $<\!z\!> = 0.5$ DLA Galaxies}

As alluded to previously, we know that DLA survey results suggest that
the incidence and cosmological mass density of low-redshift DLA systems
are larger in comparison to expectations at $z=0$ by factors of $\sim 3$
and $\sim 6.5$, respectively, although the errors in these determinations
are substantial (Rao \& Turnshek 2000 and 2001). Here we compare three
properties of the local ($z=0$) galaxy sample with the low-redshift
($<\!z\!> = 0.5$) DLA galaxy sample. First, in Figure 4 we compare
the normalized luminosity distributions of the two samples; we see
that if we ignore the corresponding information on impact parameter and
column density for the time being, these two luminosity distributions show
overall agreement.  Second, in Figure 5 we show the luminosity dependence
of how the impact parameters of the low-redshift DLA galaxies compare
with the face-on radii of local galaxies for $N_{HI} \ge 2\times10^{20}$
atoms cm$^{-2}$. There appears to be some level of agreement here as well,
after one considers the affects of imposing random orientations on the
local galaxy sample and inferring the most probable impact parameter for
a DLA system identification. We note that the probability of producing
a DLA line when $b > R_{20.3}$ is relatively small (but not zero)
for inclined galaxies. Lastly, however, we present Figure 6 and draw
attention to figure 4 in Rao \& Turnshek (2001) in these proceedings
and figures $32-34$ in Rao \& Turnshek (2000). The results shown in
these figures indicate that, to a high level of significance, the column
density distribution of low-redshift ($z<1.65$) DLA survey systems is 
markedly different from that derived for the $z=0$ galaxy sample. 
In particular, as seen in the upper panel of Figure 6, there is 
a strong relative excess of higher column density ($N_{HI} > 10^{21}$
atoms cm$^{-2}$) systems among the low-redshift ($z<1.65$) DLA systems
studied by Rao \& Turnshek (2000); however, this effect is not
readily apparent in the 
low-redshift ($<\!z\!> = 0.5$) sample of 13 DLA galaxies studied here 
(Figures $3-5$ and the lower panel of Figure 6),
which required imaging data to obtain luminosities and impact parameters.
This sample of 13 overlaps with, but is not identical to, the larger
sample studied in Rao \& Turnshek (2000), which only required neutral 
hydrogen column density information. Therefore, it should be kept in
mind that we probably have not yet studied a representative sample of 
the DLA galaxies which give rise to the highest column density systems.

\begin{figure}[h]
\plotfiddle{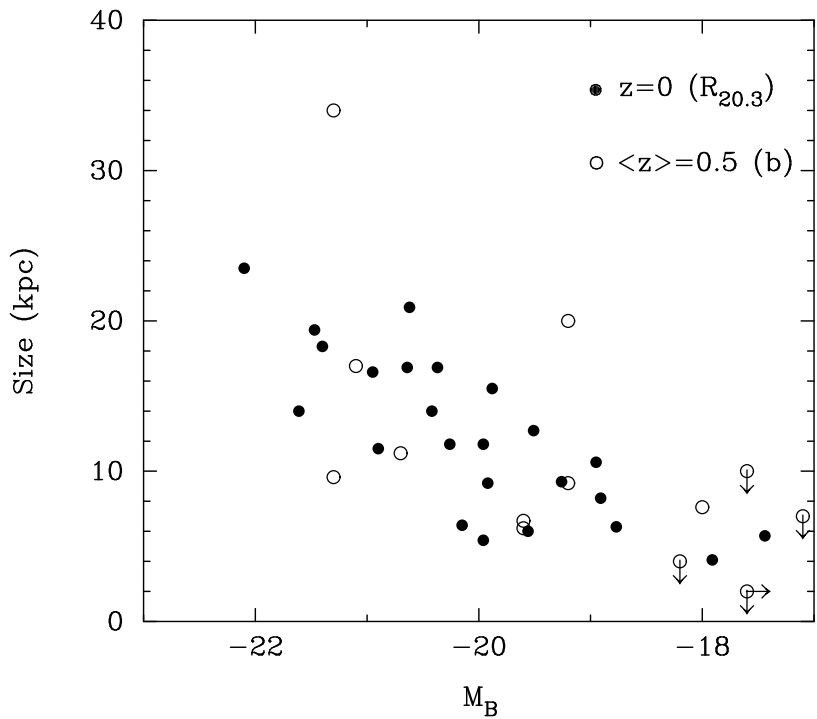}{8.7cm}{0}{75}{75}{-230}{-210}
\caption{\footnotesize The solid circles show the local ($z=0$)
galaxy results on face-on radii corresponding to $N_{HI} \ge
2\times10^{20}$ atoms cm$^{-2}$ ($R_{20.3}$) as a function of
luminosity (M$_{\rm B}$), which is the same as shown in Figure 2
(see caption). Over-plotted with open circles are the low-redshift
($<\!z\!> = 0.5$) DLA galaxy results on impact parameter
(b) as a function of luminosity (M$_B$). See the text for further
discussion.} \end{figure}

\begin{figure}[h]
\plotfiddle{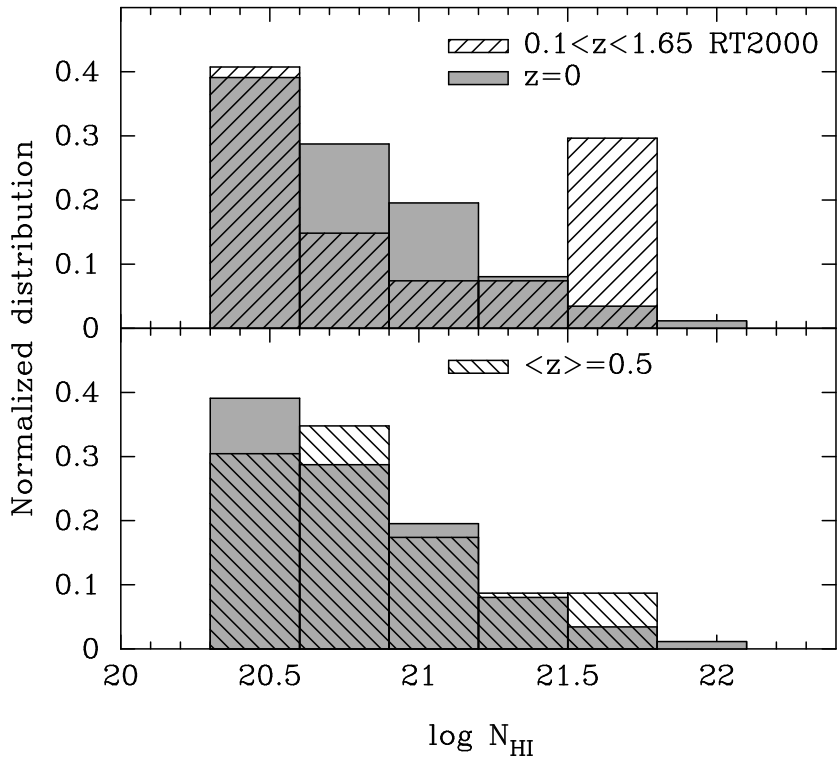}{9.7cm}{0}{80}{80}{-230}{-215}
\caption{\footnotesize Shown in grey in both panels is the equivalent
DLA HI column density distribution as derived from the local ($z=0$)
sample studied in 21 cm emission by Rao \& Briggs (1993).  For comparison
(diagonal stripes), the upper panel shows the observed low-redshift
($z<1.65$) DLA HI column density distribution from the unbiased study
of Rao \& Turnshek (2000); the lower panel shows the DLA HI column
density distribution which corresponds to the identified low-redshift
($<\!z\!> = 0.5$) DLA galaxies studied here (Figures 3$-$5).
All distributions are normalized. As the figure indicates, a 
representative number of the low-redshift ($z<1.65$) DLA galaxies
giving rise to some of the highest column density systems have not 
yet been adequately studied.} \end{figure}

\section{Conclusions and Discussion}

We have made some comparisons between the properties of local ($z=0$)
gas-rich galaxies capable of causing DLA absorption systems in the
spectra of background QSOs and the properties of the low-redshift
galaxies thought to be responsible for the DLA lines discovered in
QSO absorption-line surveys. We have referred to these galaxies as
DLA galaxies. We find that, while many of the properties of the two
samples are similar, the comparisons and overall statistical results
(Rao \& Turnshek 2000 and 2001) suggest that there may be an excess of
DLA galaxies at low redshift in the form of dwarf and/or low surface
brightness galaxies which exhibit unexpectedly large neutral hydrogen
column densities. Some examples of low-redshift DLA galaxies are shown
by Nestor et al. in these proceedings.

It would seem that there are three possible interpretations for this
finding. First, we may be missing a large fraction of dwarfs and/or
low surface brightness galaxies in local galaxy surveys that could
add to the $z=0$ DLA statistics.  At present,
there seems to be a fair amount of controversy about whether this is
possible (Schneider \& Rosenberg 2000; Zwaan, Briggs, \& Sprayberry 2001).
Also, see the separate results presented by O'Neil and Rosenberg at this
workshop.
Second, there may be substantial evolution between the $z<1.65$ QSO DLA
absorption-line samples and the $z=0$ 21 cm emission samples, however this
possibility is not easily reconciled with the notion that the DLA galaxy
population represents a relatively slowly evolving population (Pettini et
al. 1999; Rao \& Turnshek 2000). Lastly, we are still surely at the mercy
of small number statistics, and this may lead to premature speculation.

This material is based upon work supported by the National Science 
Foundation under Grant No. 9970873.

\end{document}